\documentclass[journal]{IEEEtran}
\usepackage{cite}
\usepackage{marvosym}
\usepackage{graphicx}
\usepackage{amsmath,amssymb,amsfonts}
\usepackage{algorithmic}
\usepackage{textcomp}
\usepackage{xcolor}
\usepackage{url}
\usepackage{acronym}
\usepackage{booktabs}
\usepackage{multirow}
\usepackage[hidelinks]{hyperref}
\usepackage{verbatim}
\usepackage{titlesec}
\usepackage{tikz}
 \usetikzlibrary{shapes.geometric, arrows.meta, positioning}
\usepackage{booktabs}
\usepackage{xcolor}
\newcommand{\todo}[1]{\textcolor{red}{\textbf{[TODO: #1]}}}
\titlespacing*{\section}
{0pt}{1.5ex plus .2ex minus .2ex}{0.8ex}

\titlespacing*{\subsection}
{0pt}{1ex plus .2ex minus .2ex}{0.5ex}

\titlespacing*{\subsubsection}
{0pt}{0.8ex plus .2ex minus .2ex}{0.4ex}

\begin{document}

\title{SoK: Federated Learning for Intrusion Detection
       in Vehicular Networks}

\author{
Yahya Shahsavari$^{*}$,
Reza Nourmohammadi$^{*}$,
Sara Rouhani$^{+*}$,
Kaiwen Zhang$^{*}$\\[1ex]
$^{*}$Department of Software and IT Engineering,
École de technologie supérieure (ÉTS)\\ Montréal, Québec, Canada\\
$^{+}$Department of Computer Science, University of Calgary\\ Calgary, Alberta, Canada

\ yahya.shahsavari@etsmtl.ca, 
reza.nourmohammadi.1@ens.etsmtl.ca, sara.rouhani@ucalgary.ca, kaiwen.zhang@etsmtl.ca\
}
\markboth{Cyber-AI 2026}%
{Author N. One \MakeLowercase{\textit{et al.}}: Demo Paper for Submissions to SET IJBE}

\maketitle
\begin{abstract}
Modern vehicular networks face an expanding attack surface across internal Electronic Control Units (ECUs) and external Vehicle-to-Everything (V2X) communication. Federated Learning (FL) has emerged as a decentralized paradigm to deploy Intrusion Detection Systems (IDS) without compromising data privacy. However, the vehicular FL-IDS literature suffers from fragmented methodologies and unrealistic experimental setups. This paper presents a Systematization of Knowledge (SoK) that unifies the taxonomy of vehicular attack surfaces, evaluates FL topologies, and maps adversarial threats such as poisoning and inference attacks. By auditing over 60 publications, we identify recurring pitfalls: artificial IID data splits, reliance on trivial benchmarks, weak adversarial evaluation, and omission of real-time CAN constraints. Finally, we define a forward-looking research agenda and outline minimum benchmarking requirements necessary to transition vehicular FL-IDS from optimistic simulations to secure, real-world deployment.
\end{abstract}


\section{Introduction}
\IEEEPARstart{M}{odern} vehicles are no longer isolated mechanical systems; they are deeply networked cyber-physical platforms. A contemporary automobile contains upward of 100 Electronic Control Units (ECUs) interconnected via the Controller Area Network (CAN) bus, FlexRay, Local Interconnect Network (LIN), and increasingly Automotive Ethernet~\cite{rajapaksha2023ai}. Beyond the vehicle boundary, Vehicle-to-Everything (V2X) communication standards (i.e., Dedicated Short-Range Communications (DSRC) and Cellular V2X (C-V2X)) enable interaction with roadside infrastructure, other vehicles, pedestrians, and network services. These connectivity layers expand the attack surface dramatically: remote compromise of in-vehicle systems can alter braking, steering, or acceleration; misbehaving V2X messages can corrupt shared situational awareness and trigger accidents~\cite{hasan2020securing}.

Traditional Intrusion Detection Systems for vehicular networks face a structural tension: effective anomaly detection is data-hungry and benefits from cross-fleet knowledge~\cite{zhang2018selecting}, yet vehicle traffic data is operationally sensitive, legally constrained by data-protection regulations, such as GDPR and UN Regulation (No.~155 - Cyber security and cyber security management system), and practically difficult to centralize due to bandwidth and latency constraints~\cite{alsamiri2023federated}. Federated Learning (FL)~\cite{mcmahan2017communication} resolves this tension in principle: vehicles train local models on-board and exchange only model updates (gradients or parameters) with an aggregation server, without exposing raw traffic.


FL-based vehicular IDS research has grown rapidly since 2020, but the literature remains fragmented across protected layers, FL topologies, model choices, datasets, adversarial assumptions, and evaluation metrics. This SoK examines which findings are well supported, which may result from unrealistic assumptions, and which challenges remain open. This SoK paper makes the following contributions:
\begin{itemize}
  \item A unified taxonomy of vehicular network architectures, attack surfaces, and threat actors, serving as a common reference frame (\ref{sec:background}).
  \item A systematic classification of FL-IDS architectures applied to vehicular networks, covering aggregation strategies, model families, and federation topologies (\ref{sec:fl_ids}).
  \item A structured analysis of adversarial threats targeting FL-IDS in vehicular settings (i.e., data poisoning, model poisoning, Byzantine attacks, and inference attacks), and a critique of existing defences (\ref{sec:adversarial}).
  \item A comprehensive evaluation of datasets, simulation tools, and evaluation methodologies used in the literature, revealing pervasive evaluation pitfalls (\ref{sec:evaluation}).
  \item A gap analysis and forward-looking agenda (\ref{sec:open_problems}).
\end{itemize}

\textbf{Scope:} In this paper we focus on studies that apply FL, including horizontal, vertical, and federated transfer learning variants, to intrusion or anomaly detection in vehicular environments (in-vehicle, V2X/VANET, or both). We consider works published between 2017 and early 2026. We exclude general FL and vehicular-security surveys, as well as IDS approaches that do not employ FL.

\section{Survey Methodology}
\label{sec:methodology}

To ensure transparent and auditable coverage, we followed a structured literature identification, screening, and coding protocol. Given the multi-year scope of this SoK (2017--2026) and the fast-moving nature of the vehicular FL-IDS literature, coverage was built iteratively rather than through a single, logged database query pass: we combined keyword search with citation snowballing, re-screened the corpus as new work appeared, and applied a fixed coding schema to every included study. We report this process as an iterative literature audit rather than a formal PRISMA systematic review~\cite{page2021prisma}. We did not capture per-database hit counts (e.g., records identified, duplicates removed, and records screened) at each iteration. Furthermore, a single-pass, fully logged search was not feasible. This is because we assembled and re-verified the corpus across multiple search sessions over an extended period. Below, we describe our exact steps so that the process (if not the exact hit counts) remains transparent. Exact intermediate counts are reproducible.

\subsection{Search Strategy}
\label{subsec:search-strategy}
 
We queried four databases: IEEE Xplore, ACM Digital Library, Scopus,
and arXiv (restricted to the cs.CR, cs.NI, and cs.LG categories),
using Boolean combinations of the following term groups:
 
\begin{itemize}
    \item \textbf{FL terms}: \{``federated learning''\}
    \item \textbf{IDS terms}: \{``intrusion detection'', ``misbehavior
    detection'', ``misbehaviour detection'', ``anomaly detection''\}
    \item \textbf{Domain terms}: \{``vehicular'', ``VANET'', ``V2X'',
    ``CAN bus'', ``automotive'', ``IoV''\}
\end{itemize}
 
\noindent The three groups were combined conjunctively (FL terms)
AND (IDS terms) AND (domain terms), with database-specific field
restrictions (title/abstract/keywords where supported). Scopus was
included specifically for its cross-publisher aggregation, which
sweeps in Springer, Elsevier, and MDPI venues alongside
IEEE/ACM-indexed work; we did not additionally query SpringerLink
directly, so Springer-published proceedings not indexed by Scopus at
query time may be under-represented. The search window covered
January 2017 -- February 2026, reflecting the publication of the
original FedAvg algorithm \cite{mcmahan2017communication} as the field's
natural starting point. Searches were repeated periodically
throughout the drafting process (rather than once) to capture newly
published work, and were supplemented by backward and forward
citation snowballing from an initial seed set of highly relevant
papers and adjacent surveys. 

\subsection{Inclusion and Exclusion Criteria}
\label{subsec:inclusion-exclusion}
 
\textbf{Inclusion criteria (all must hold):}
\begin{enumerate}
    \item[\textbf{I1}] Peer-reviewed publication or arXiv preprint.
    \item[\textbf{I2}] Proposes, evaluates, or surveys a
    federated-learning-based (horizontal, vertical, or federated
    transfer learning) IDS or MDS.
    \item[\textbf{I3}] Applied to CAN, V2X, or hybrid vehicular
    settings.
    \item[\textbf{I4}] Published between January 2017 and February
    2026.
    \item[\textbf{I5}] Written in English.
\end{enumerate}
 
\textbf{Exclusion criteria (any one is sufficient to exclude):}
\begin{enumerate}
    \item[\textbf{E1}] General-purpose FL surveys with no vehicular
    focus.
    \item[\textbf{E2}] Vehicular IDS papers that do not employ FL.
    \item[\textbf{E3}] Duplicate or extended-abstract versions of an
    already-included full paper.
    \item[\textbf{E4}] Papers without any empirical evaluation
    (purely conceptual/position papers).
\end{enumerate}
 
\noindent Applying these criteria across the iterative search
process described above yielded a corpus of more than 60 studies
directly examining FL-based intrusion or misbehaviour detection in
vehicular settings, in addition to the adjacent surveys, attack
taxonomies, dataset papers, and FL algorithmic works cited for
background and comparison throughout this SoK. We report this figure
as an approximate corpus size rather than an exact, independently
reproducible count (see Section~\ref{subsec:threats-validity}).

\subsection{Data Extraction and Coding Protocol}
\label{subsec:coding-protocol}

Each of the $+60$ included studies was coded against a fixed extraction schema (See Table~\ref{tab:coding-schema}). Coding was performed by the authors during the iterative review process described above; papers added in later search passes were coded against the same schema to keep the extracted fields consistent across the full corpus.


\begin{table}[t]
\centering
\caption{Data extraction schema applied to all included studies}
\label{tab:coding-schema}
\begin{tabular}{p{2.1cm}p{5.6cm}}
\toprule
\textbf{Field} & \textbf{Coded values} \\
\midrule
Dataset(s) & Car-Hacking, OTIDS, ROAD, SynCAN, CAN-T\&T, VeReMi,
VeReMi Ext., other/custom \\
Partitioning & IID (random/uniform), non-IID by vehicle identity,
non-IID by other criterion, not reported \\
Byzantine eval. & Yes / No / Not reported \\
Byzantine fraction & Numeric \% if reported, else N/A \\
Aggregation rule & FedAvg, FedProx, Krum, trimmed mean/median,
FLTrust, personalized (FedPer/pFedMe), other \\
Privacy mechanism & None, DP, SecAgg, HE, combination \\
Topology & Centralized, hierarchical, decentralized/gossip,
asynchronous \\
Inference latency reported & Yes / No \\
Model family & LSTM/BiLSTM, GRU, CNN, autoencoder, GNN, transformer,
MLP/DNN, ensemble \\
\bottomrule
\end{tabular}
\end{table}

\subsection{Threats to Validity}
\label{subsec:threats-validity}

As with any systematic literature study, several threats to validity
apply: 

\begin{itemize}
    \item \textit{Publication bias}: papers reporting negative or weak results are underrepresented in the venues we searched, which may inflate the apparent maturity of the field. 
    \item \textit{Search completeness}: our Boolean query, while broad, may miss papers using non-standard terminology (e.g., ``collaborative learning'' instead of ``federated learning''); we partially mitigate this via backward snowballing from included papers' reference lists.
    \item \textit{Single/small-team
coding bias}: the primary data extraction (Section~\ref{subsec:coding-protocol}) was performed by authors without independent third-party verification. Hence, systematic misclassification cannot be fully ruled out. 

    \item \textit{Language and access bias}: we restricted inclusion to English-language, electronically accessible publications, which may exclude relevant work published in other venues or languages.
\end{itemize}

\color{black}
\section{Related Works}

Several surveys are adjacent to this SoK. Agrawal et al.~\cite{agrawal2022federated} survey FL-based IDSs across multiple domains, including IoT, healthcare, and vehicular environments, but do not systematically distinguish between in-vehicle and V2X settings. Alsamiri and Alsubhi~\cite{alsamiri2023federated} provide a taxonomy of FL-based IDSs for the Internet of Vehicles (IoV), but do not examine adversarial threats or assess the quality and realism of the experimental evidence. Belenguer et al.~\cite{belenguer2025review} present a systematic literature review of FL-based IDSs across multiple domains; however, vehicular environments receive only limited coverage.
Outside the FL literature, Rajapaksha et al.~\cite{rajapaksha2023ai} survey AI-based IDSs for in-vehicle networks, while Lampe et al.~\cite{lampe2023survey} review deep learning approaches for automotive intrusion detection. Neither work considers FL.

This SoK differs from prior surveys and reviews by: (a)~systematically assessing the evidence supporting claims in the literature; (b)~explicitly addressing both intra-vehicle and V2X FL-based IDSs; (c)~analyzing adversarial threats specific to vehicular FL; and (d)~identifying recurring evaluation shortcomings and corresponding research opportunities.

\section{Background}\label{sec:background}

\subsection{Vehicular Network Architectures}

Vehicular networks span two distinct domains whose security requirements and FL deployment constraints differ substantially.

\subsubsection{Intra-Vehicle Networks}

The dominant intra-vehicle communication standard remains the Controller Area Network (CAN), originally developed by Bosch in 1986~\cite{avatefipour2018state}. CAN is a broadcast, multi-master bus designed for reliability rather than security; it provides no source authentication, encryption, or protection against node impersonation~\cite{aliwa2021cyberattacks}. Consequently, an Electronic Control Unit (ECU) that gains physical or remote access to the bus can inject arbitrary messages. Although CAN Flexible Data-rate (CAN FD) and Automotive Ethernet (100BASE-T1/1000BASE-T1) are increasingly adopted in modern vehicles, they do not inherently provide comprehensive security mechanisms.

CAN arbitration IDs, payloads, and timing patterns form the main features used by anomaly-based IDSs. Because CAN traffic is safety-critical and time-sensitive, IDS inference must operate within strict real-time constraints.~\cite{wu2019survey, ozdemir2024survey}.

\subsubsection{Inter-Vehicle and V2X}

V2X includes V2V, V2I, V2P, and V2N communication through DSRC or C-V2X technologies, enabling frequent exchange of safety-critical messages such as BSMs and CAMs.

A central security threat in V2X environments is \emph{misbehaviour}, where participants broadcast false position, speed, heading, or other contextual information. Such behaviour may arise from malicious actions— Sybil attacks, replay attacks, and data falsification, or from sensor failures and measurement errors~\cite{kamel2020simulation}. Misbehaviour Detection Systems (MDSs) serve as the V2X counterpart to traditional intrusion detection systems.

\subsubsection{Emerging Architectures}

MEC nodes, RSUs, and cloud backends create a multi-tier hierarchy that naturally supports hierarchical FL, with training and aggregation distributed across vehicles, edge nodes, and cloud servers~\cite{liu2021vehicular}.

\subsection{Attack Taxonomy}

Table~\ref{tab:attack_taxonomy} presents our unified attack taxonomy spanning both intra-vehicle and inter-vehicle environments. The taxonomy organizes threats across three distinct layers: the intra-vehicle CAN bus, the inter-vehicle V2X communication plane, and the FL aggregation pipeline. Each layer exposes a different attack surface and requires tailored countermeasures.

\subsubsection{Intra-Vehicle Attacks (CAN Bus)}
The CAN bus is the dominant in-vehicle network protocol, connecting ECU responsible for braking, steering, and engine management. Because CAN was designed for reliability rather than security (lacking source authentication or encryption) it is inherently vulnerable to a range of injection-based attacks~\cite{koscher2010experimental,bozdal2018survey,miller2015remote}.

\textbf{DoS/Flooding:} An attacker with physical or remote bus access repeatedly transmits high-priority frames (e.g., \texttt{0x000}), monopolizing bandwidth and preventing legitimate ECU communications. This can disrupt critical safety functions such as ABS and throttle control~\cite{verma2024comprehensive}.

\textbf{Fuzzy Attack:} The adversary injects frames with randomly chosen identifiers and payloads. Although individually these frames may be ignored, their cumulative effect disrupts ECU state machines, potentially causing erratic behaviour such as unintended acceleration or dashboard failures~\cite{lee2017otids}.

\textbf{Replay Attack:} Captured legitimate frames are retransmitted at a later time or at a higher rate. Because CAN carries no timestamps or sequence numbers, receiving ECU cannot distinguish replayed messages from fresh ones. This can lead to stale or contradictory control state~\cite{bozdal2018survey,checkoway2011comprehensive}.

\textbf{Spoofing/Impersonation:} The attacker injects frames bearing a victim ECU's identifier, overriding or supplementing its output. Since CAN arbitration is deterministic, a sufficiently fast injector can silence the legitimate node entirely and assume full control of the associated vehicle function~\cite{iehira2018spoofing}.

\subsubsection{Inter-Vehicle Attacks (V2X)}

\begin{table}[t]
\centering
\caption{Unified Attack Taxonomy for Vehicular Networks}
\label{tab:attack_taxonomy}
\renewcommand{\arraystretch}{1.2}
\begin{tabular}{p{1.6cm} p{1.5cm} p{1.8cm} p{1.8cm}}
\toprule
\textbf{Layer} & \textbf{Attack Type} & \textbf{Mechanism} & \textbf{Impact} \\
\midrule
\multirow{4}{*}{Intra (CAN)} 
  & DoS/Flooding & Flood bus with ID=0x000 & Bus unavailability \\
  & Fuzzy & Random ID injection & Erratic ECU behaviour \\
  & Replay & Re-transmit captured frames & State confusion \\
  & Spoofing/ Impersonation & Inject with victim's ID & Safety control bypass \\
\midrule
\multirow{4}{*}{Inter (V2X)} 
  & Sybil & Multiple fake identities & Map pollution \\
  & GPS Spoofing & False positional data & Wrong routing \\
  & Replay & Retransmit old BSMs & Stale awareness \\
  & Data Falsification & Manipulate reported state & Accident provocation \\
\midrule
\multirow{3}{*}{FL-Specific}
  & Data Poisoning & Corrupt local training data & Degraded global model \\
  & Model Poisoning & Manipulate local updates & Backdoor/DoS \\
  & Inference & Reconstruct training data & Privacy breach \\
\bottomrule
\end{tabular}
\end{table}

V2X communications (DSRC and C-V2X) enable vehicles to share safety-critical information such as position, speed, and road hazards via BSM. The broadcast-and-trust model of these protocols makes them susceptible to identity-based and data-integrity attacks~\cite{sedar2023comprehensive,kenney2011dedicated,marojevic2018c}.

\textbf{Sybil Attack:} A malicious vehicle fabricates multiple pseudonymous identities, each broadcasting plausible but fictitious traffic data. Aggregated by neighbours and infrastructure, these phantom nodes can create virtual traffic jams, ghost accidents, or false road hazard warnings, polluting digital maps and misdirecting vehicles~\cite{adele2024survey,malathi2014detection}.

\textbf{GPS Spoofing:} The adversary transmits counterfeit GNSS signals that override authentic satellite signals at the victim's receiver. The victim vehicle consequently reports and navigates by an incorrect position, potentially causing wrong routing decisions or collisions in platooning and cooperative driving scenarios~\cite{dasgupta2022sensor}.

\textbf{Replay Attack:} Previously recorded BSM are retransmitted after a delay. Neighbours receiving these stale messages form an outdated awareness picture: vehicles may react to hazards that have already resolved or fail to detect new ones, degrading cooperative perception~\cite{alnasser2019cyber}.

\textbf{Data Falsification/False Data Injection:} Rather than fabricating identities, the attacker manipulates the payload of its own legitimately authenticated messages—reporting false speed, heading, or event severity. Such attacks are difficult to detect with classical cryptographic measures because the message is signed by a genuine certificate; detection requires plausibility checks or trust management overlays~\cite{hasan2020securing}. 

\subsubsection*{FL-Specific Attacks}

Federated learning is increasingly proposed for collaborative intrusion detection and driver-behaviour modelling in vehicular networks, as it avoids centralizing raw  data~\cite{mcmahan2017communication}. However, the distributed training process introduces a new attack surface at the model-update layer~\cite{li2025threats}.

\textbf{Data Poisoning:} A compromised vehicle deliberately corrupts its local training dataset (mislabeling normal frames as attacks, or vice versa) before local training begins. The tainted gradients propagate to the global model after aggregation, degrading detection accuracy for targeted classes~\cite{tolpegin2020data}.

\textbf{Model Poisoning:} Instead of corrupting data, the adversary directly manipulates the local model update (gradient or weight delta) before submission. Sophisticated variants craft updates that cause the aggregated global model to embed a hidden backdoor—correctly classifying benign inputs while misclassifying attacker-chosen trigger patterns or simply perform a Byzantine denial-of-service on model convergence~\cite{bagdasaryan2020backdoor}.

\textbf{Inference/Membership Inference Attack:} By querying the shared global model or observing intermediate updates, an adversary can reconstruct statistical properties of participants' private training data, infer whether a specific record was used in training, or even recover raw samples via gradient inversion. These attacks violate the privacy guarantees that motivate federated learning in the first place~\cite{liu2022membership}.

\subsection{Federated Learning Fundamentals}
FL~\cite{mcmahan2017communication} is a distributed learning paradigm where clients $\mathcal{K}$ collaboratively train a global model $\mathbf{w}$ by minimizing:
\[
  \min_{\mathbf{w}} \sum_{k \in \mathcal{K}} \frac{n_k}{n} \mathcal{L}_k(\mathbf{w}),
\]
where $n_k$ is client $k$'s data size, $n = \sum_k n_k$, and $\mathcal{L}_k$ is client $k$'s local loss. The canonical algorithm FedAvg~\cite{mcmahan2017communication} performs $E$ local Stochastic Gradient Descent (SGD) steps per round before aggregating weighted averages of client parameters at the server.

In vehicular Federated Learning (FL), clients are vehicles/fleets, servers are cloud/RSU/MEC, and communication is intermittent with limited bandwidth. A key overlooked challenge is the highly non-IID data due to diverse traffic, attacks, and driving patterns.

\section{FL-IDS Architectures in Vehicular Networks}\label{sec:fl_ids}
Vehicular networks face vulnerabilities in both intra-vehicle buses and V2X links, so deploying Federated Learning for Intrusion Detection Systems (FL-IDS) requires tight co-design of ML pipelines and automotive hardware constraints. This section systematically categorizes the vehicular FL-IDS landscape across operational topologies, core ML model families, global aggregation strategies, and privacy-enhancing mechanisms.
\subsection{Federation Topology}
The physical and logical topology of the FL system determines latency, communication overhead, fault tolerance, and the threat model for Byzantine participants.

\subsubsection{Centralized (Single-Tier) FL}
The simplest topology: all vehicles communicate directly with a central aggregation server (e.g., an Original Equipment Manufacturer (OEM) cloud or a dedicated RSU cluster). FedAvg or one of its variants aggregates updates every round. Works in this category include early FL-IDS proposals for IoV~\cite{hbaieb2022federated} and CAN-bus FL-IDS exploiting Bidirectional Long Short-Term Memory (BiLSTM) models~\cite{driss2022federated}. The principal limitation is the single point of failure at the aggregator, vulnerability of the aggregator to inference attacks, and inability to serve highly mobile vehicles that drift in and out of connectivity.

\subsubsection{Hierarchical (Multi-Tier) FL}
A two or three-tier hierarchy is better suited to vehicular mobility. Vehicles aggregate locally with nearby RSUs or MEC nodes (edge tier), which in turn aggregate with a cloud server (global tier). This dramatically reduces communication latency and allows edge nodes to maintain a regional model that adapts to local traffic conditions. Hierarchical FL-IDS proposals have demonstrated improved convergence in high-mobility settings~\cite{althunayyan2024robust} and naturally partition the non-IID problem by geographic region.

\subsubsection{Decentralized/Gossip FL}
This topology eliminates the single point of failure but introduces consensus challenges, is more susceptible to Byzantine participants, and requires careful aggregation to prevent gradient divergence. SD-VANET-based FL-IDS proposals~\cite{sathishkumar2026federated} use SDN controllers as lightweight coordinators in lieu of a full aggregation server.

\subsubsection{Asynchronous FL}
Standard FedAvg is synchronous: every vehicle must complete its local round before global aggregation. In VANETs with high churn (vehicles joining and leaving the network), synchronous FL is impractical. Asynchronous variants where the server aggregates whenever a sufficient subset of updates arrives are increasingly explored for vehicular settings~\cite{nie2025mobility}, at the cost of increased staleness.

\subsection{Model Families}

Table~\ref{tab:model_summary} summarises the ML model families employed in FL-IDS for vehicular networks. LSTM-based models dominate the in-vehicle FL-IDS literature due to their ability to capture temporal patterns in CAN message sequences~\cite{hossain2020lstm}. The BiLSTM variant used in~\cite{driss2022federated} processes sequences in both directions, capturing forward and backward temporal context. For V2X misbehaviour detection, graph neural networks exploit the vehicular communication graph topology~\cite{yuce2024misbehavior,ibrahim2026spatiotemporal}.

\begin{table}[t]
\centering
\caption{Model Families Used in Vehicular FL-IDS}
\label{tab:model_summary}
\renewcommand{\arraystretch}{1.2}
\begin{tabular}{p{1.8cm} p{1.5cm} p{2.0cm} p{1.5cm}}
\toprule
\textbf{Model} & \textbf{Target} & \textbf{Strengths} & \textbf{Limitations} \\
\midrule
LSTM/Bi-LSTM & CAN & Temporal dependencies & High memory \\
GRU & CAN & Faster than LSTM & Less expressive \\
CNN & CAN/V2X & Spatial patterns & Fixed-length input \\
Autoencoder & CAN/V2X & Unsupervised & Threshold sensitivity \\
GNN & V2X & Graph structure & Complex topology \\
Transformer & CAN/V2X & Long-range context & High compute cost \\
MLP/DNN & Both & Fast inference & Weaker temporal modelling \\
Ensemble & Both & Diversity & Aggregation overhead \\
\bottomrule
\end{tabular}
\end{table}

\subsection{Aggregation Strategies}

The aggregation function determines how client updates are combined into the global model. Beyond FedAvg, several aggregation strategies appear in the vehicular FL-IDS literature:

\begin{itemize}
  \item \textbf{FedProx}~\cite{li2020federated}: Adds a proximal term to client objectives, improving convergence under heterogeneous data. Particularly relevant in vehicular settings where non-IID data induces gradient divergence.

  \item \textbf{Krum}~\cite{blanchard2017machine}: Selects the update closest (in Euclidean distance) to its $k$ nearest neighbours, providing Byzantine resilience. Computationally expensive for large models.

   \item \textbf{Trimmed Mean / Coordinate-wise Median}: Removes extreme values in each coordinate dimension before averaging. Effective against untargeted poisoning but susceptible to sophisticated attacks~\cite{fang2020local}.

   \item \textbf{FLTrust}~\cite{cao2020fltrust}: The server maintains a small clean dataset and uses cosine similarity between the server gradient and client gradients to assign trust scores. Well-suited to vehicular settings where OEMs can maintain a small labelled reference dataset.

   \item \textbf{Personalized Aggregation (FedPer, pFedMe)}: Allows each vehicle to maintain a personalized local model while benefiting from a shared global representation. Addresses the challenge that each vehicle has a unique traffic fingerprint~\cite{arivazhagan2019federated,t2020personalized}.
\end{itemize}

\subsection{Privacy-Preserving Mechanisms}
FL alone does not guarantee privacy, as gradients may reveal training data through model inversion and membership inference attacks~\cite{lyu2020threats}. Three complementary privacy mechanisms appear in the vehicular FL-IDS literature:

\begin{itemize}
  \item \textbf{Differential Privacy (DP)}: Gaussian or Laplace noise is added to gradients before upload. DP provides formal privacy guarantees but degrades detection accuracy, especially in non-IID settings with small local datasets~\cite{abadi2016deep}.
  \item \textbf{Secure Aggregation (SecAgg)}: Cryptographic masking (e.g., Scryptographic masking and secret-sharing techniques) ensures the server only sees the aggregate, not individual updates. Incurs significant computational overhead on resource-constrained vehicular ECUs~\cite{bonawitz2017practical}.
  \item \textbf{Homomorphic Encryption (HE)}: Allows the server to aggregate encrypted updates. provides strong confidentiality for model updates but is currently impractical for large models due to computational cost~\cite{aono2017privacy}.
\end{itemize}

The DRL-PBFL scheme~\cite{pan2025privacy} combines deep reinforcement learning for Byzantine robustness with a novel secure aggregation protocol to simultaneously address both poisoning and inference threats in vehicular FL. This work is notable for jointly optimizing robustness and privacy as a combination rarely addressed together.

\section{Adversarial Threats to Vehicular FL-IDS}\label{sec:adversarial}
Although FL protects raw data privacy, its distributed nature introduces a large attack surface across highly mobile and physically accessible vehicles. This section examines the vehicular FL-IDS threat landscape, focusing on Byzantine behavior, data and model poisoning, backdoor attacks, and client-side inference vulnerabilities.

\subsection{Threat Model}

A vehicular FL-IDS adversary may be \emph{external} (attacker has no foothold inside the FL system) or \emph{internal} (attacker controls one or more FL participants). Internal adversaries are Byzantine: they can deviate arbitrarily from the FL protocol. We additionally distinguish between \emph{integrity} goals (degrading the global model) and \emph{confidentiality} goals (extracting private training data).

The threat landscape is complicated by the vehicular context. A malicious vehicle can be a deliberately compromised fleet vehicle, a remotely exploited OBD-II port, or a Sybil identity created by a single physical attacker. Unlike FL for healthcare or finance, vehicular FL nodes are mobile, transient, and may not be under continuous OEM monitoring.

\subsection{Data Poisoning}
In data poisoning, the adversary corrupts the local training dataset of compromised vehicles before or during FL training. Label-flipping attacks systematically mislabel attack traffic as benign, inducing a model that fails to detect specific intrusion types. Against CAN-bus FL-IDS, a poisoned vehicle can inject synthetic CAN messages into its local dataset, corrupting the learned temporal patterns. The work of~\cite{mansourian2025enhancing} demonstrates that adversarial evasion attacks (which alter CAN frames to avoid detection) can be adapted to data poisoning with limited perturbation, and that standard FL aggregation (FedAvg) provides no protection.

\subsection{Model Poisoning}
Model poisoning, where the adversary directly manipulates the local model update before submission, is substantially more powerful than data poisoning~\cite{fang2020local}. The attacker crafts model parameters that maximize the global model's test error while remaining within the detection radius of robust aggregators (e.g., Krum). For vehicular FL-IDS, the high mobility and frequent client turnover make it difficult to distinguish malicious outlier updates from legitimate updates generated by vehicles in unusual traffic conditions.

Benchmarking studies on autonomous vehicle datasets~\cite{almutairi2024comprehensive} have found that: (a)~IID data distribution is \emph{not} universally more secure than non-IID, contradicting a common assumption; (b)~client-selection strategies in cross-device settings can serve as a simple yet effective first line of defence; and (c)~state-of-the-art Byzantine-robust aggregators fail against adaptive attacks that exploit knowledge of the aggregation.

\subsection{Backdoor Attacks}
A backdoor attack causes the global model to behave normally on clean inputs while misclassifying a specific \emph{triggered} input as benign. In vehicular CAN networks, a trigger may consist of a particular arbitration ID and payload pattern. Since triggers are rare and overall accuracy remains high, backdoors are difficult to detect. Vehicular FL studies have only recently begun addressing backdoor attacks~\cite{demir2025targeted}, and no existing vehicular FL-IDS proposal provides formal guarantees against them.

\subsection{Inference Attacks}
Even without malicious updates, FL participants can attempt to reconstruct training data from the global model or from model updates observed during aggregation. Gradient inversion attacks~\cite{geiping2020inverting} have shown that full CAN frame sequences can be approximately reconstructed from a single gradient update, undermining the primary privacy motivation for FL. Membership inference attacks can reveal whether a specific vehicle's traffic appeared in training, exposing driving patterns. DRL-PBFL~\cite{pan2025privacy} explicitly defends against gradient inversion by honest-but-curious MEC nodes, a threat overlooked by most vehicular FL-IDS proposals.

\subsection{Sybil Attacks on FL Participants}
A Sybil attacker creates multiple fake vehicle identities to amplify the influence of malicious updates during aggregation. In open V2X networks where vehicle identities are pseudonymous, Sybil creation is low-cost. Standard FL aggregation rules that weight updates by data volume or participant count are acutely vulnerable. Trust management systems based on reputation scores or PKI-anchored certificates~\cite{liu2024lightweight} partially mitigate this threat but introduce reliance on a trusted authority, which may itself be attacked or unavailable.

\subsection{Summary: Defence Coverage}

Table~\ref{tab:defenses} summarises defensive mechanisms against the adversarial threats described above and the extent to which each is addressed in the vehicular FL-IDS literature.

\begin{table}[t]
\centering
\caption{Adversarial Threat Coverage in Vehicular FL-IDS Literature}
\label{tab:defenses}
\renewcommand{\arraystretch}{1.2}
\begin{tabular}{p{2.0cm} p{1.4cm} p{1.4cm} p{1.8cm}}
\toprule
\textbf{Threat} & \textbf{Defence Type} & \textbf{Coverage} & \textbf{Gap} \\
\midrule
Data poisoning & Robust aggregation. & Partial & Adaptive attacks \\
Model poisoning & Krum/TMean & Partial & Adaptive attacks \\
Backdoor & Detection & Minimal & Formal guarantees \\
Gradient inversion & SecAgg/DP & Partial & Overhead vs.\ accuracy \\
Sybil & Reputation & Partial & Trust anchor needed \\
Byzantine + inference & DRL-PBFL & Single work & Scalability unknown \\
\bottomrule
\end{tabular}
\end{table}

\begin{figure}[!t]
\centering
\includegraphics[
width=\columnwidth,
keepaspectratio,
trim={0 0 0 0},
clip
]{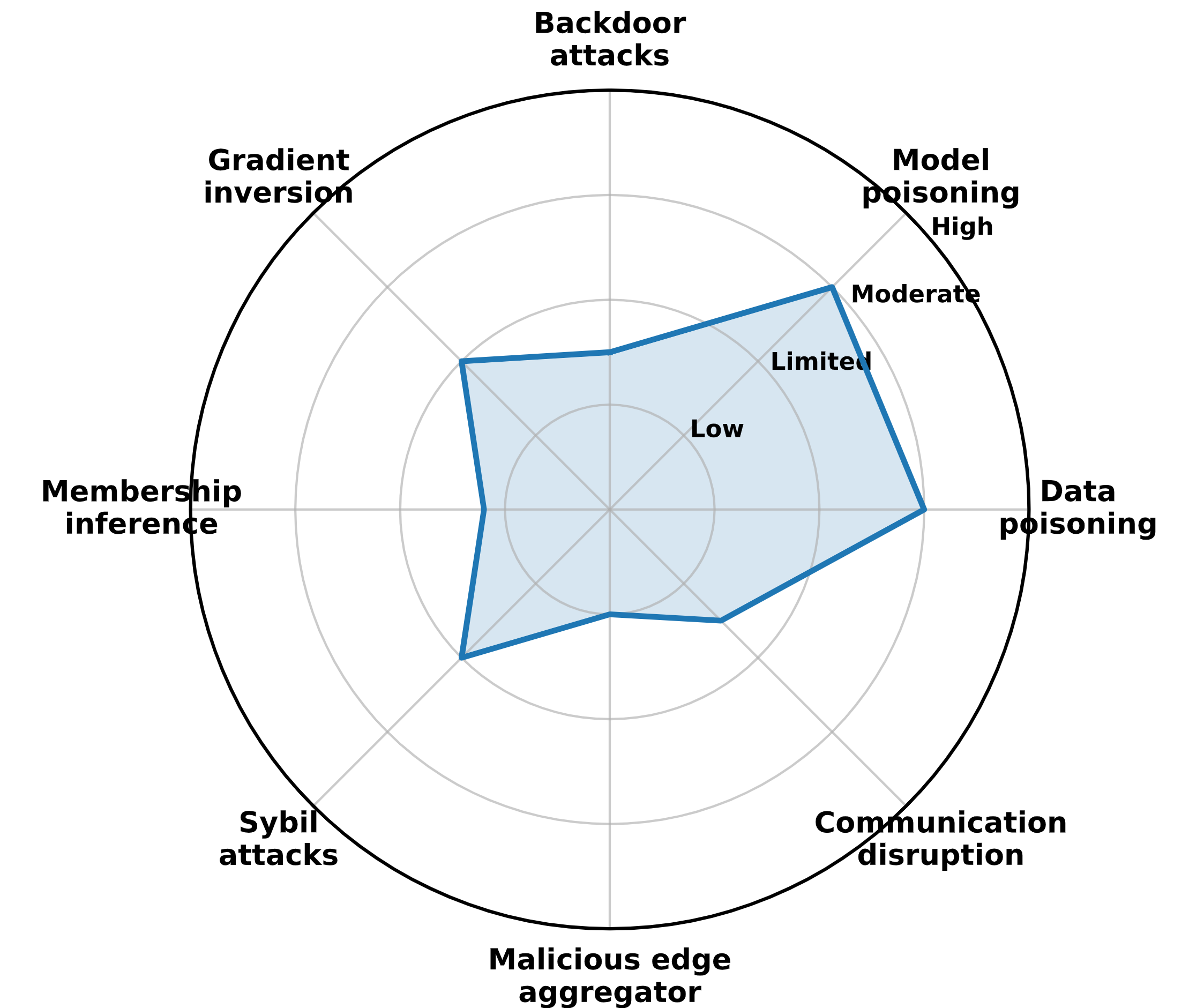}
\caption{Threat coverage in vehicular FL-IDS}
\label{fig:threat-coverage-fl-ids}
\end{figure}


\section{Datasets and Evaluation}\label{sec:evaluation}
Validating the efficacy of a vehicular FL-IDS relies entirely on the quality and realism of the underlying empirical benchmarks. This section evaluates the principal intra-vehicle and V2X datasets used in the literature, critically uncovering pervasive methodology flaws regarding data partitioning, adversarial assumptions, and real-time execution constraints.
\subsection{Intra-Vehicle Datasets}

\subsubsection{Car-Hacking Dataset (HCRL)}
Released by the Hacking and Countermeasure Research Lab (HCRL) at Korea University~\cite{song2020vehicle}, this dataset contains 500~seconds of benign CAN traffic and four attack types (i.e. DoS, fuzzy, and two spoofing variants (RPM and gear)). It is the most widely used CAN-IDS benchmark, but has significant limitations for FL research: all attack frames alter ID frequencies dramatically (F1-scores above 99\% are routine~\cite{al2024can}; benign data was collected while the car was stationary; and there is no natural FL partition aligned with vehicle identity.

\subsubsection{OTIDS Dataset (HCRL)}
The OTIDS dataset~\cite{lee2017otids} captures DoS, fuzzy, and impersonation attacks on a KIA Soul using remote CAN frames. Unusually, it lacks ground-truth labels as attributes; instead, attack injection intervals are documented, which multiple works have found to be inaccurate. It is the only publicly available CAN dataset featuring remote frames, but its labelling limitations render it unreliable for supervised learning.

\subsubsection{ROAD Dataset (ORNL)}
The Real ORNL Automotive Dynamometer (ROAD) dataset~\cite{verma2020road} is currently regarded as the most realistic CAN intrusion dataset, featuring physically verified injection attacks including correlated signal fabrication and targeted ID masquerade. It reflects actual vehicle dynamics and provides a more faithful test of IDS performance. Its use in FL-IDS literature is growing but not yet dominant.

\subsubsection{SynCAN}
SynCAN~\cite{hanselmann2020canet} is a synthetic dataset with ground truth designed for evaluating methods that rely on signal semantics rather than frame statistics. It enables controlled experiments but lacks the noise and artifacts of real CAN traffic.

\subsubsection{CAN-Train-and-Test}
A curated multi-vehicle dataset~\cite{lampe2024can} explicitly designed to support cross-vehicle generalization studies. It includes benign data from multiple vehicle makes and several attack types, making it better suited to evaluating FL models that must generalize across vehicle fleets.

\subsection{Inter-Vehicle / V2X Datasets}
\subsubsection{VeReMi / VeReMi Extension}
The Vehicular Reference Misbehaviour (VeReMi) dataset~\cite{kamel2020simulation} and its extension are the standard benchmarks for V2X misbehaviour detection. Generated using the LuST traffic scenario, Veins, and SUMO, VeReMi contains 225 simulations at varying vehicle densities and attacker ratios. Attack types include random position offset, constant offset, eventual stop, and Sybil. The FL literature is beginning to use VeReMi for federated misbehaviour detection~\cite{huang2024semi}, though most works use pre-existing train/test splits that are not designed for realistic FL client partitioning.

\subsection{Evaluation Methodology Critique}
Our analysis of over 60 FL-IDS papers for vehicular networks reveals the following pervasive evaluation pitfalls:

\subsubsection{IID Data Partitioning}
\textbf{Finding:} A substantial majority of surveyed works randomly and uniformly split a single dataset across FL clients, creating artificially IID local distributions~\cite{belenguer2025review}. This is fundamentally inconsistent with the vehicular setting, where vehicles encounter different conditions, attackers, and attack frequencies.

\subsubsection{Single-Dataset Evaluation}
\textbf{Finding:} The majority of works evaluate on a single dataset, most commonly the Car-Hacking dataset. As noted above, this dataset's attacks are trivially detectable by frequency-based methods, meaning reported F1-scores above 99\% convey limited information about real-world performance.

\subsubsection{Absence of Byzantine Participants}
\textbf{Finding:} A minority of vehicular FL-IDS proposals evaluate in the presence of Byzantine (malicious or faulty) FL participants. Given that a compromised vehicle can be an FL participant, this omission is critical.

\subsubsection{No Real-Time Constraint Evaluation}
\textbf{Finding:} To the best of our knowledge, no surveyed paper measures inference latency against the CAN bus timing budget (frame period 0.5--100~ms depending on ID priority). A model that achieves 99.9\% accuracy but requires 500~ms per inference provides zero practical utility on CAN.

\subsubsection{Simulation Toolchain Mismatch}
\textbf{Finding:} V2X FL-IDS papers using SUMO/Veins~\cite{lopez2018microscopic} simulate mobility and communication but rarely model real 802.11p or C-V2X PHY/MAC characteristics. Communication drops, handover delays, and channel congestion, which directly affect FL convergence, are typically omitted.

\subsection{Dataset and Toolchain Summary}

Table~\ref{tab:datasets} summarises the principal datasets used in vehicular FL-IDS research.

\begin{table}[t]
\centering
\caption{Vehicular IDS Datasets and Their FL Suitability}
\label{tab:datasets}
\renewcommand{\arraystretch}{1.2}
\begin{tabular}{p{1.7cm} p{0.8cm} p{0.8cm} p{1.4cm} p{1.5cm}}
\toprule
\textbf{Dataset} & \textbf{Layer} & \textbf{Real?} & \textbf{FL Partition?} & \textbf{Limitation} \\
\midrule
Car-Hacking & CAN & Yes & No & Trivial attacks \\
OTIDS & CAN & Yes & No & Unreliable labels \\
ROAD & CAN & Yes & No & Limited attacks \\
SynCAN & CAN & Synth. & Yes & Synthetic \\
CAN-T\&T & CAN & Yes & Partial & Multi-vehicle \\
VeReMi & V2X & Simul. & No & Simulated only \\
VeReMi Ext. & V2X & Simul. & No & Simulated only \\
\bottomrule
\end{tabular}
\end{table}

\begin{figure}[!t]
\centering
\includegraphics[
width=\columnwidth,
trim={0 0 0 0},
clip
]{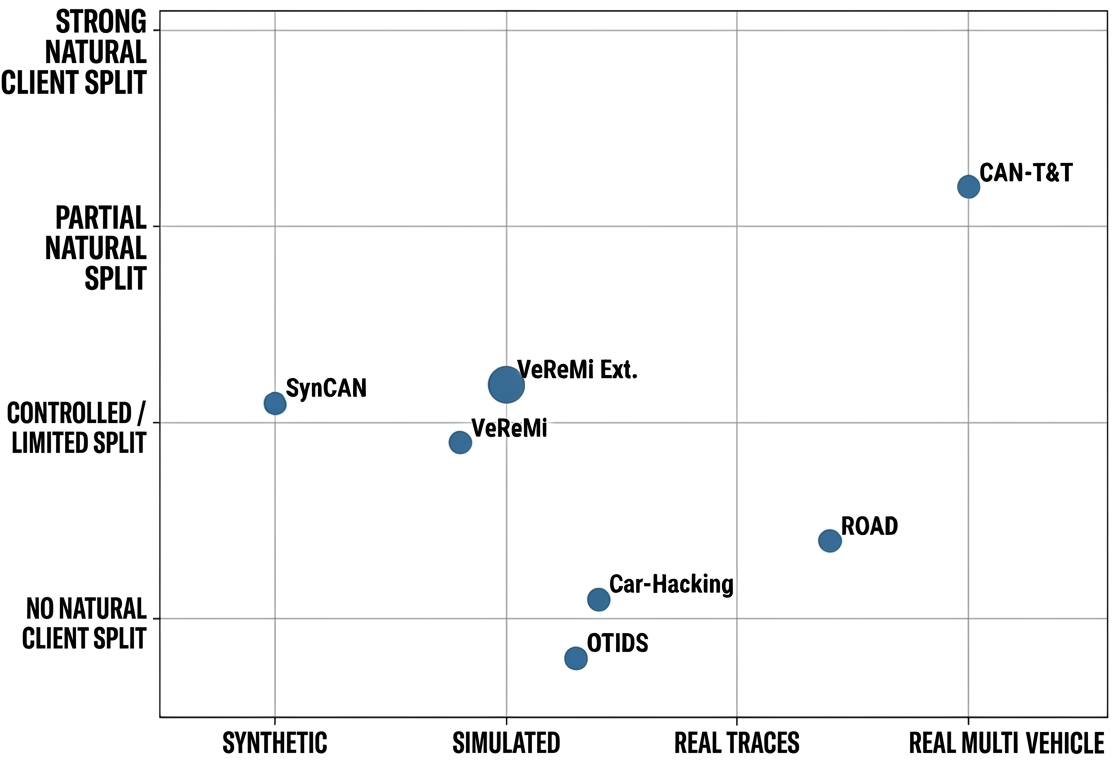}
\caption{Dataset realism and FL suitability in vehicular IDS evaluation}
\label{fig:dataset-realism-fl-suitability}
\end{figure}

\section{Systematization: What Have We Established?}\label{sec:systematization}

Drawing on the preceding analysis, we now systematize the principal claims in the vehicular FL-IDS literature and assess the strength of evidence for each.

\subsection{Claim 1: FL preserves privacy in vehicular IDS}

\textbf{Stated:} Numerous papers claim FL is "privacy-preserving" because raw data never leaves the vehicle.

\textbf{Assessment: PARTIALLY SUPPORTED.} FL prevents direct data sharing, but gradient inversion attacks~\cite{geiping2020inverting} demonstrate that CAN frame sequences can be approximately reconstructed from gradient updates. Only works that combine FL with SecAgg or DP provide meaningful privacy guarantees. The majority of vehicular FL-IDS proposals do not employ these mechanisms, making the privacy claim misleading.

\subsection{Claim 2: FL-IDS achieves competitive detection accuracy}

\textbf{Stated:} FL-IDS proposals consistently report F1-scores $\geq 0.95$ on standard benchmarks.

\textbf{Assessment: CONDITIONALLY SUPPORTED.} High reported accuracy is a consequence of evaluation on the Car-Hacking dataset, whose attacks trivially alter ID frequencies. On more realistic datasets (ROAD, CAN-Train-and-Test) and with non-IID partitioning, accuracy drops substantially. The community lacks a shared non-IID benchmark, making cross-paper comparison unreliable.

\subsection{Claim 3: FL reduces communication overhead compared to centralized learning}

\textbf{Stated:} By sharing model updates rather than raw data, FL reduces bandwidth requirements.

\textbf{Assessment: SUPPORTED WITH CAVEATS.} This claim holds for raw data volume. However, in vehicular settings with intermittent connectivity, FL synchronization rounds impose \emph{latency} overhead that may be more critical than bandwidth. Asynchronous FL variants reduce latency but introduce staleness. No work has conducted a rigorous communication cost analysis under realistic V2X channel conditions.

\subsection{Claim 4: Hierarchical FL is superior for vehicular settings}

\textbf{Stated:} Multi-tier FL with RSU/MEC edge aggregation improves convergence under mobility.

\textbf{Assessment: PLAUSIBLE BUT UNDERVALIDATED.} Hierarchical FL-IDS proposals show improvements in simulated settings, but simulations typically model neither real channel impairments nor adversarial participants at the edge tier. A compromised RSU is a high-impact single point of failure in a hierarchical topology — a threat not addressed in any reviewed paper.

\subsection{Claim 5: Robust aggregation defends against Byzantine vehicles}

\textbf{Stated:} Algorithms such as Krum, trimmed mean, and FLTrust protect FL-IDS from malicious participants.

\textbf{Assessment: WEAKLY SUPPORTED.} Benchmarking studies~\cite{fang2020local,almutairi2024comprehensive} demonstrate that adaptive attackers who optimize against a known aggregation rule can bypass all current Byzantine-robust aggregators. In vehicular deployments, attackers may have access to the aggregation algorithm (it may be standardized).

\section{Open Problems, Future Directions and Research Agenda}\label{sec:open_problems}

Although FL for automotive security is advancing quickly, significant structural, algorithmic, and regulatory barriers still block real-world deployment. This section maps these open challenges and outlines a concrete research agenda to achieve resilient, standardized, and production-ready vehicular FL-IDS.

\subsection{Realistic Evaluation Benchmarks}

The field urgently needs a standardized FL-IDS benchmark for vehicular networks that: (a)~provides realistic non-IID partitions based on vehicle identity; (b)~includes both CAN and V2X scenarios; (c)~incorporates Byzantine participants; (d)~specifies inference latency requirements; and (e)~defines reproducible evaluation protocols. We propose the following minimum requirements for a credible FL-IDS evaluation:

\begin{enumerate}
  \item Use at least two datasets from different vehicle platforms.
  \item Partition data by vehicle identity or by geographic region (not randomly).
  \item Evaluate with at least 10\% Byzantine participants.
  \item Report inference latency alongside accuracy metrics.
  \item Release code and partitioning details for reproducibility.
\end{enumerate}

\subsection{Non-IID Data and Personalization}
The non-IID problem in vehicular FL is more severe than in standard FL benchmarks because vehicles are mobile, encounter different attackers, and have different ECU configurations. Personalized FL approaches (e.g., FedPer, pFedMe, Ditto) are promising but have not been systematically evaluated in the vehicular IDS context.

\subsection{Edge Aggregator Security}
Hierarchical vehicular FL uses RSUs and MEC nodes as intermediate aggregators. These infrastructure nodes may be physically compromised or remotely exploited. However, the risks posed by a malicious edge aggregator that tampers with regional models before forwarding them to the cloud remain largely unexplored in vehicular IDSs.

\subsection{Real-Time Constraints}
CAN-bus IDSs require hard real-time constraints. FL adds computational overhead to ECUs, but co-designing FL schedules within ECU resource budgets without disrupting safety-critical CAN processing remains an open challenge.

\subsection{Cross-OEM Federation}
Effective fleet-level threat intelligence needs FL across different OEMs, but cross-OEM federation faces major challenges: heterogeneous CAN formats, proprietary ECUs, competitive reluctance to share models, and regulatory liability. No existing FL-IDS work addresses this; all assume single-OEM settings.

\subsection{V2X Misbehaviour Detection at Scale}
Existing federated misbehaviour detection proposals for V2X rely on VeReMi, which simulates only a few hundred vehicles in a Luxembourg city scenario. Real urban deployments will involve thousands of vehicles with rapid topology changes. The scalability of federated MDS — in convergence speed, communication cost, and Byzantine resilience — remains unexplored at realistic urban scales.

\subsection{Integration with Automotive Standards}

Vehicle cybersecurity is heavily regulated by UN R155 (mandating CSMS for European type-approved vehicles) and ISO/SAE 21434. FL-IDS systems must integrate into these frameworks, but key open issues remain: who certifies the global FL model, how model updates are audited, and who bears liability if a poisoned model fails to detect an attack.

\subsection{LLMs and Foundation Models for Vehicular IDS}
Foundation models pre-trained on large CAN or V2X corpora may help vehicular FL-IDS reduce training rounds and improve generalization to unseen attacks. Future work should evaluate their communication overhead, edge feasibility, privacy leakage, and backdoor risks during federated fine-tuning.

\section{Conclusion}\label{sec:conclusion}
FL is a promising approach for vehicular intrusion detection, but current evidence is weakened by unrealistic evaluations, including IID data splits, simple attack datasets, missing Byzantine participants, and absent real-time latency analysis. As a result, claims about FL’s privacy, accuracy, and Byzantine robustness remain only conditionally supported.

We call on the community to adopt the minimum evaluation standards proposed in Section ~\ref{sec:open_problems}, develop and share non-IID vehicular FL benchmarks, and engage
with the regulatory landscape shaped by UN R155 and ISO/SAE 21434. The safety-critical nature of vehicular networks demands that the security guarantees of FL-IDS be grounded in rigorous, adversarially-aware evaluation — not optimistic experimental conditions that will not survive contact with real deployments.

\def\IEEEbibitemsep{0pt plus .1pt}
\small

\bibliographystyle{IEEEtran}
\bibliography{References} 

\end{document}